\documentclass[prb,preprint]{revtex4}
\usepackage{graphicx}
\begin{document}
\draft
\begin{abstract}
Starting from a model of an elastic medium, we derive equations of
motion that are identical in form to Dirac's equation for a spin 1/2
particle with mass, coupled to electromagnetic and gravitational
interactions. The mass and electromagnetic terms are not added by
hand but emerge naturally from the formalism.  A two dimensional
version of this equation is derived by starting with a model in
three dimensions and deriving equations for the dynamics of the
lowest Fourier modes assuming one dimension to be periodic.
Generalizations to higher dimensions are discussed.
\end{abstract}
\pacs{PACS numbers: } \vspace{.5in}
\title{A Derivation of Dirac's Equation From a Model of an Elastic Medium}
\author{John Baker}
\affiliation{ 26970 Hayward Blvd., Apt 507; Hayward, CA 94542}
\maketitle

\section{Introduction} Dirac's equation describes the behavior of
particles with mass and spin and how they couple to the
electromagnetic field.  The usual form of Dirac's equation is

\[ (\imath\gamma^{\mu}\partial_{\mu}-m)\Psi(x)=0
\]

The electromagnetic field is introduced by the minimal coupling
prescription\cite{ref:Peskin} $\partial_{\mu}\rightarrow D_{\mu}$,
with

\[D_{\mu}=\partial_{\mu}+\imath  A_{\mu}(x)
\]

where $A_{\mu}$ is the electromagnetic vector potential.  Dirac's
equation can be further coupled to gravity (at the classical level)
using the prescription\cite{ref:Brill_Wheeler}
\[
\partial_\mu\rightarrow\partial_\mu-\Gamma_\mu
\]
and the equation then takes the
form\cite{ref:Finster,ref:Brill_Wheeler,ref:Brill_Cohen,ref:Smoller_Finster,ref:Smoller_Finster2}
\begin{equation}
\label{eq:full_dirac}
\tilde{\gamma}^{\mu}[\imath\partial_{\mu}-\imath\Gamma_{\mu}-A_\mu]\Psi(x)-m\Psi(x)=0
\end{equation}
where $\Gamma_\mu$ is known as the spin connection, $A_\mu$ is the
electromagnetic vector potential, and $m$ is the mass. The
gravitational coupling enters through the modified dirac matrices
$\tilde{\gamma}_{\mu}$ which satisfy the anticommutation relation

\[\{\tilde{\gamma}^{\mu},\tilde{\gamma}^{\nu}\}=I g^{\mu\nu}.
\]
  and the operator $\tilde{\gamma}^{\mu}[\partial_{\mu}-\Gamma_{\mu}]$
is (in the absence of electromagnetic interactions) the covariant
derivative for spinor fields in a curved
space\cite{ref:Brill_Wheeler}.

The above form of Dirac's equation describes the dynamics of the
spinor field $\Psi$ when coupled to the scalar fields $A_{\mu}$ and
gravity.  There are two additional equations which describe the
dynamics of $A_{\mu}$ and $g_{\mu\nu}$, these are the Einstein field
equations

\begin{equation}
\label{eq:einstein_field}
 R_{\mu\nu}-\frac{1}{2}R=T_{\mu\nu}
\end{equation}

and Maxwell's equations

\begin{equation}
\label{eq:maxwell}
 \nabla_\mu F^{\mu\nu}=4\pi
e\bar{\Psi}\gamma^\nu\Psi
\end{equation}

The Equations~(\ref{eq:full_dirac}), (\ref{eq:einstein_field}) and
(\ref{eq:maxwell}) are collectively known as the
Einstein-Dirac-Maxwell
equations\cite{ref:Smoller_Finster,ref:Smoller_Finster2,ref:Krori}.
The subject of this paper is Equation~(\ref{eq:full_dirac}).  We
will show that the equations of motion of an elastic solid have the
same form as Equation~(\ref{eq:full_dirac}) with the mass end
electromagnetic term emerging naturally from the formalism.

\section{Elasticity Theory}
\label{sec:elasticity_theory}
 The theory of elasticity is usually
concerned with the infinitesimal deformations of an elastic
body\cite{ref:Love,ref:Sokolnikoff,ref:Landau_Lifshitz,ref:Green_Zerna,ref:Novozhilov}.
We assume that the material points of a body are continuous and can
be assigned a unique label $\vec{a}$. For definiteness the elastic
body can be taken to be a three dimensional object so each point of
the body may be labeled with three coordinate numbers $a_{i}$ with
$i=1,2,3$.

If this three dimensional elastic body is placed in a large ambient
three dimensional space then the material coordinates $a_{i}$ can be
described by their positions in the 3-D fixed space coordinates
$x_{i}$ with $i=1,2,3$.  In this description the material points
$a_{i}(x_1,x_2,x_3)$ are functions of $\vec{x}$. A deformation of
the elastic body results in infinitesimal displacements of these
material points. If before deformation a material point $a_0$ is
located at fixed space coordinates $x_1,x_2,x_3$ then after
deformation it will be located at some other coordinate
$x'_1,x'_2,x'_3$.  The deformation of the medium is characterized at
each point by the displacement vector
\[u_i=x'_i-x_i
\]
which measures the displacement of each point in the body after
deformation.

It is the aim of this paper to take this model of an elastic medium
and derive from it equations of motion that have the same form as
Dirac's equation.
%In doing so we have to distinguish between the
%intrinsic coordinates of our medium which we will call "internal"
%coordinates and the fixed space coordinates which facilitates our
%derivation of the equations of motion.
%
%If in the absence of any deformation we were to form a coordinate
%system consisting of three orthogonal vectors
%
%\[
%\vec{e_i}=\frac{\partial}{\partial a_i}
%\]
% at any point in space then these vectors
%would define the notions of direction inside the medium.  In the
%undeformed state we may take the external coordinates to coincide
%with the material coordinates $\partial/\partial
%a_i=\partial/\partial x_i$. If there were a distortion of the
%elastic medium these coordinates would no longer coincide and our
%original vector would no longer be orthogonal with reference to the
%fixed space coordinates.
%
%The approach that we will use in this paper is to derive equations
%of motion using the fixed space coordinates and then translate this
%to the internal coordinates of our space. The justification for
%translating to the internal coordinates is that later in this paper
%we will assume that one of the internal coordinates is periodic and
%therefore the Fourier transform of the equations of motion require
%that internal coordinates be used. Therefore any dynamics that are
%expressed in terms of the fixed space vectors
%$\frac{\partial}{\partial x_i}$, need to be translated into dynamics
%that are appropriate to the internal vectors
%$\frac{\partial}{\partial a_i}$.

We first consider the effect of a deformation on the measurement of
distance. After our elastic body is deformed, the distances between
its points changes as measured with the fixed space coordinates. If
two points which are very close together are separated by a radius
vector $dx_i$ before deformation, these same two points are
separated by a vector $dx'_i=dx_i+du_i$. The square distance between
the points before deformation is then $ds^2=dx_1^2+dx_2^2+dx_3^2$.
Since these coincide with the material points in the undeformed
state, this can be written $ds^2=da_1^2+da_2^2+da_3^2$. The squared
distance after deformation can be written\cite{ref:Landau_Lifshitz}
$ds'^{2}=dx_1'^2+dx_2'^2+dx_3'^2=\sum_i
dx_i'^2=\sum_i(da_i+du_i)^2$. The differential element $du_i$ can be
written as $du_i=\sum_i\frac{\partial u_i}{\partial a_k }da_k$,
which gives for the distance between the points

\begin{eqnarray*}
ds'^2&=&\sum_i\left(da_i + \sum_k\frac{\partial u_i}{\partial
a_k}da_k\right) \left(da_i + \sum_l\frac{\partial u_i}{\partial
a_l}da_l\right)\\
&=&\sum_i\left(da_i da_i + \sum_k\frac{\partial u_i}{\partial
a_k}da_i da_k+ \sum_l\frac{\partial u_i}{\partial a_l}da_i da_l +
\sum_k\sum_l\frac{\partial u_i}{\partial a_k}\frac{\partial
u_i}{\partial a_l}\right)\\
&=&\sum_i\sum_k\left(\delta_{ik}+\left(\frac{\partial u_i}{\partial
a_k}+\frac{\partial u_k}{\partial a_i}\right)+\sum_l\frac{\partial
u_l}{\partial
a_i}\frac{\partial u_l}{\partial a_k}\right) da_k da_l\\
&=&\sum_{ik}\left(\delta_{ik}+2\epsilon'_{ik}\right)da_i da_k
\end{eqnarray*}
 where $\epsilon'_{ik}$ is
\begin{equation}
\label{eq:strain_tensor}
\epsilon'_{ik}=\frac{1}{2}\left(\frac{\partial u_i}{\partial
a_k}+\frac{\partial u_k}{\partial a_i}+\sum_l \frac{\partial
u_l}{\partial a_i}\frac{\partial u_l}{\partial a_k}\right)
\end{equation}
 The
quantity $\epsilon'_{ik}$ is known as the strain tensor. It is
fundamental in the theory of elasticity. In most treatments of
elasticity it is assumed that the displacements $u_i$ as well as
their derivatives are infinitesimal so the last term in
Equation~(\ref{eq:strain_tensor}) is dropped.  This is an
approximation that we will not make in this derivation.

The quantity
\begin{eqnarray}
\label{eq:metric}
 g_{ik}&=&\delta_{i,k}+\frac{\partial u_i}{\partial
a_k}+\frac{\partial u_k}{\partial a_i}+\sum_l \frac{\partial
u_l}{\partial a_i}\frac{\partial u_l}{\partial a_k}\\
&=&\delta_{i,k}+2\epsilon'_{ik}\nonumber
\end{eqnarray}
 is the metric for our system and
determines the distance between any two points.

That this metric is simply the result of a coordinate transformation
from the flat space metric can be seen by writing the metric in the
form\cite{ref:Millman_Parker}
\[
 g_{\mu\nu}=
 \left( \begin{array}{lll}{\displaystyle
\frac{\partial x'_1}{\partial a_1}}& {\displaystyle\frac{\partial
x'_2}{\partial
a_1}} & {\displaystyle\frac{\partial x'_3}{\partial a_1}}\\[15pt]
 {\displaystyle\frac{\partial x'_1}{\partial a_2}}& {\displaystyle\frac{\partial x'_2}{\partial a_2}} &
{\displaystyle\frac{\partial x'_3}{\partial a_2}}\\[15pt]
{\displaystyle\frac{\partial x'_1}{\partial a_3}}&
{\displaystyle\frac{\partial x'_2}{\partial a_3}} &
{\displaystyle\frac{\partial x'_3}{\partial a_3}}
\end{array}
 \right)
 \left(\begin{array}{lll}
 {\displaystyle 1}& {\displaystyle 0} & {\displaystyle 0}\\[15pt]
 {\displaystyle 0}& {\displaystyle 1}& {\displaystyle 0}\\[15 pt]
 {\displaystyle 0}& {\displaystyle 0} & {\displaystyle 1}
 \end{array}
 \right)
 \left(\begin{array}{lll}
{\displaystyle \frac{\partial x'_1}{\partial a_1}}&
{\displaystyle\frac{\partial x'_1}{\partial
a_2}} & {\displaystyle \frac{\partial x'_1}{\partial a_3}}\\[15pt]
 {\displaystyle\frac{\partial x'_2}{\partial a_1}}& {\displaystyle\frac{\partial x'_2}{\partial a_2}} &
{\displaystyle\frac{\partial x'_2}{\partial a_3}}\\[15pt]
{\displaystyle\frac{\partial x'_3}{\partial a_1}}&
{\displaystyle\frac{\partial x'_3}{\partial a_2}} &
{\displaystyle\frac{\partial x'_3}{\partial a_3}}
\end{array}
\right)
\]
\[
=J^TIJ
\]
where
\[
\frac{\partial x'_\mu}{\partial
x_\nu}=\delta_{\mu\nu}+\frac{\partial u_\mu}{\partial a_\nu}.
\]
and $J$ is the Jacobian of the transformation. Later in section
\ref{sec:fourier_transform} we will show that the metric for the
Fourier modes of our system is not a simple coordinate
transformation.

The inverse matrix $(g^{ik})=(g_{ik})^{-1}$ is given by
$(g^{ik})=(J^{-1})(J^{-1})^T$ where
\begin{equation}
J^{-1}=\left(\begin{array}{lll} {\displaystyle\frac{\partial
a_1}{\partial x'_1}}& {\displaystyle\frac{\partial a_1}{\partial
x'_2}} & {\displaystyle\frac{\partial a_1}{\partial x'_3}}\\[15pt]
 {\displaystyle\frac{\partial a_2}{\partial x'_1}}& {\displaystyle\frac{\partial a_2}{\partial x'_2}} &
{\displaystyle\frac{\partial a_2}{\partial x'_3}}\\[15pt]
{\displaystyle\frac{\partial a_3}{\partial x'_1}}&
{\displaystyle\frac{\partial a_3}{\partial x'_2}} &
{\displaystyle\frac{\partial a_3}{\partial x'_3}}
\end{array}
\right)
\end{equation}
 This yields for the inverse metric
 \begin{eqnarray}
 g^{ik}&=&\delta_{ik}-\frac{\partial
u_i}{\partial x_k}-\frac{\partial u_k}{\partial x_i}+\sum_l
\frac{\partial u_l}{\partial x_i}\frac{\partial u_l}{\partial x_k}\\
&=&\delta_{ik}-2\epsilon_{ik}\nonumber
 \end{eqnarray}
 where $\epsilon_{ik}$ is defined by
 \[
 \epsilon_{ik}=\frac{1}{2}\left(\frac{\partial
u_i}{\partial x_k}+\frac{\partial u_k}{\partial x_i}-\sum_l
\frac{\partial u_l}{\partial x_i}\frac{\partial u_l}{\partial
x_k}\right)
\]
We see that the metric components involves derivatives of the
displacement vector with respect to the internal coordinates and the
inverse metric involves derivatives with respect to the fixed space
coordinates.

\section{Equations of Motion}
\label{sec:EOM}
 In the following we will use the notation
\[
u_{\mu\nu}=\frac{\partial u_\mu}{\partial x_\nu}
\]
 and therefore the inverse strain tensor is
 \[
 \epsilon_{\mu\nu}=\frac{1}{2}\left(u_{\mu\nu}+u_{\nu\mu}+\sum_\beta
 u_{\beta \mu}u_{\beta\nu}\right).
 \]

We will use the lagrangian method to derive the equations of motion
for our system. Our model consists of an elastic solid embedded in a
$3$ dimensional euclidean space.
%Our
%approach will be to derive the equations of motion in the fixed
%space coordinates and transform them to equations in the internal
%coordinates
% by using the
%displacement vector $u_i=a_i-x_i$ which allows us to write the
%partial derivative with respect to the $x_i$ coordinates as,
%
%\begin{eqnarray*}
%\label{eq:derivative_transformation} \frac{\partial}{\partial
%x_i}&=&\sum_j\frac{\partial a_j}{\partial
%x_i}\frac{\partial}{\partial a_j}\\
%&=& \sum_j\left(\frac{\partial
%x_j}{\partial x_i}+\frac{\partial u_j}{\partial x_i}\right)\frac{\partial}{\partial a_j}\\
%&=& \sum_j\left(\delta_{ij}+\frac{\partial u_j}{\partial x_i}\right)\frac{\partial}{\partial a_j}\\
%\end{eqnarray*}

In the following we work in the fixed space coordinates and take the
strain energy as the lagrangian density of our system. This approach
leads to the usual equations of equilibrium in elasticity
theory\cite{ref:Love,ref:Novozhilov}. The strain energy is quadratic
in the strain tensor $\epsilon^{\mu\nu}$ and can be written as
\[
E=\sum_{\mu \nu\alpha\rho} C_{\mu \nu\alpha\rho}\, \epsilon_{\mu\nu}
\epsilon_{\alpha\rho}
\]

The quantities $C_{\mu \nu\alpha\rho}$ are known as the elastic
stiffness constants of the material\cite{ref:Sokolnikoff}.  For an
isotropic space most of the coefficients are zero and in $3$
dimensions, the lagrangian density reduces to
\begin{equation}
\label{eq:lagrangian_3D}
 L=(\lambda +
2\mu)\left[\epsilon_{11}^2+\epsilon_{22}^2+\epsilon_{33}^2\right] +
2 \lambda \left[\epsilon_{11} \epsilon_{22}+ \epsilon_{11}
\epsilon_{33} + \epsilon_{22}\epsilon_{33}\right] + 4\mu
\left[\epsilon_{12}^2 + \epsilon_{13}^2 + \epsilon_{33}^2\right]
\end{equation}
where $\lambda$ and $\mu$ are known as Lam\'e
constants\cite{ref:Sokolnikoff}.
%This lagrangian would be
%appropriate for an elastic solid in equilibrium (ie
%$\partial/\partial t=0$).

The usual Lagrange equations,
\[
\sum_\nu\frac{d}{dx_\nu}\left(\frac{\partial L}{\partial u_{\rho
\nu}}\right) - \frac{\partial L}{\partial u_\rho}=0,
\]
apply with each component of the displacement vector treated as an
independent field variable. Since our Lagrangian contains no terms
in the field $u_\rho$, Lagrange's equations reduce to
\[
\sum_\nu\frac{d}{dx_\nu}\left(\frac{\partial L}{\partial u_{\rho
\nu}}\right)=0.
\]
The quantity
\[
V_\rho=\sum_\nu\frac{d}{dx_\nu}\left(\frac{\partial L}{\partial
u_{\rho \nu}}\right)
\]
is a vector and as such can always be written as the sum of the
gradient of a scalar and the curl of a vector or
\[
\vec{V}=\nabla\phi+\nabla\times \vec{A}.
\]

From this decomposition we can immediately conclude,
\begin{equation}
\label{eq:Laplaces_equation}
 \nabla^2 \phi=0
\end{equation}
We see therefore that the scalar quantity $\phi$ in the medium obeys
Laplace's equation.
\subsection{Physical Interpretation of $\phi$}

To understand the physical origin of $\phi$ we derive its form in
the usual infinitesimal theory of elasticity. The advantage of the
infinitesimal theory is that an explicit form of the vector
$\vec{V}$ may be obtained.  In the infinitesimal theory of
elasticity the strain components $u_{\mu\nu}$ are assumed to be
small quantities and therefore the quadratic terms in the strain
tensor are dropped and the strain tensor reduces
to\cite{ref:Landau_Lifshitz}
\[
\epsilon_{\mu\nu}=\frac{1}{2}\left(u_{\mu\nu} + u_{\nu\mu}\right)
\]

Using the above Lagrangian we obtain the explicit form

\[
V_\rho=\sum_\nu\frac{d}{dx_\nu}\left(\frac{\partial L}{\partial
u_{\rho \nu}}\right)=(2\mu+2\lambda)\frac{\partial \sigma}{\partial
x_\rho} + 2\mu \nabla^2 u_\rho=0,
\]
where $\sigma=u_{11}+u_{22}+u_{33}\equiv \nabla\cdot\vec{u}$.

Finally taking the divergence of $\vec{V}$ yields
\[
\nabla^2\sigma=0
\]
From this we see that the scalar in the infinitesimal theory is the
divergence of the strain field $\sigma=\nabla \cdot \vec{u}$.  It is
an invariant with respect to change of coordinates and in general
varies from point to point in the medium. This exercise exhibits the
physical origin of $\phi$ which to lowest order in the strain
components is the divergence of the strain field.

In this work however will not make the infinitesimal approximation
and we will work with the scalar $\phi$ and not $\sigma$. In most of
what follows, the exact form of $\phi$ is not important. It is only
important that such a quantity exists and obeys Laplace's equation.

%\subsection{Infinitesimal Equations}
%To connect this derivation to the usual infinitesimal theory of
%elasticity we write out the Lagrange equations as follows.  Denote
%the derivative of the Lagrangian with respect to the strain
%components as
%\[
%U_{\mu\nu}=\frac{\partial L}{\partial u_{\mu\nu}}.
%\]
%The above form of the lagrangian yields
%\[
%U_{\rho\nu}=2\lambda\sigma_1\frac{\partial a_\rho}{\partial
%x_\nu}+4\mu\sum_\alpha\frac{\partial a_\rho}{\partial
%x_\alpha}\epsilon_{\alpha\nu}
%\]
%where $\sigma_1=\epsilon_{11}+\epsilon_{22}+\epsilon_{33}$ and
%\[
%\frac{\partial a_\rho}{\partial
%x_\nu}=\delta_{\rho\nu}+\frac{\partial u_\rho}{\partial x_\nu}.
%\]
%
%In the infinitesimal approximation all terms that have products of
%$u_{\rho\nu}$ are ignored and only the terms linear in $u_{\rho\nu}$
%are kept.  This results in
%\[
%U_{\rho}=2\lambda \frac{\partial \sigma}{\partial
%x_\rho}+4\mu\nabla^2 u_\rho
%\] where $\sigma=u_{11}+u_{22}+u_{33}$.   Applying the gradient $\vec{\nabla}$ to
%the vector $\vec{U}$ yields
%\[
%2\lambda\,\nabla^2\sigma + 4\mu \nabla^2\sigma=0
%\]
%or
%\[
% \nabla^2 \sigma=0
%\]
%From this we see that the scalar in the infinitesimal theory is the
%divergence of the strain field $\sigma=\nabla \cdot \vec{u}$. It is
%an invariant with respect to change of coordinates and in general
%varies from point to point in the medium.  Since we do not make the
%infinitesimal approximation in this paper we will work with the
%scalar $\phi$ and not $\sigma$.

\subsection{Internal Coordinates}
The central results of this work will be given in sections
\ref{sec:internal_coordinates} and \ref{sec:fourier_transform},
where we will take one of our internal coordinates to be periodic
and we will Fourier transform all quantities in that coordinate. We
therefore need to translate the equations of motion $\nabla^2\phi=0$
from the fixed space coordinates to the internal coordinates.  For
clarity, in the remainder of this text we change notation slightly
and write the internal coordinates not as $a_i$ but as $x_i'$ and
the fixed space coordinates will be unprimed and denoted $x_i$. Now
using $u_i=x_i'-x_i$ we can write

\begin{eqnarray}
\label{eq:coordinate_change} \frac{\partial}{\partial
x_i}&=&\sum_j\frac{\partial x'_j}{\partial
x_i}\frac{\partial}{\partial x'_j} \nonumber \\
&=& \sum_j\left(\frac{\partial
x_j}{\partial x_i}+\frac{\partial u_j}{\partial x_i}\right)\frac{\partial}{\partial x'_j}\nonumber\\
&=& \sum_j\left(\delta_{ij}+\frac{\partial u_j}{\partial
x_i}\right)\frac{\partial}{\partial x'_j}
\end{eqnarray}
%\begin{equation}
%\label{eq:coordinate_change} \frac{\partial}{\partial
%x_i}=\frac{\partial}{\partial x'_i}+\sum_j\frac{\partial
%u_j}{\partial x_i}\frac{\partial}{\partial x'_j}
%\end{equation}

Equation~(\ref{eq:coordinate_change}) relates derivatives in the
fixed space coordinates $x_i$ to derivatives in the material
coordinates $x'_i$. As mentioned earlier, in the standard treatment
of elastic solids the displacements $u_i$ as well as their
derivatives are assumed to be infinitesimal and so the second term
in Equation~(\ref{eq:coordinate_change}) is dropped and there is no
distinction made between the $x_i$ and the $x'_i$ coordinates.  In
this paper we will keep the nonlinear terms in
Equation~(\ref{eq:coordinate_change}) when changing coordinates.
Hence we will make a distinction between the two sets of coordinates
and this will be pivotal in the derivations to follow.

We will now demonstrate that Laplace's equation
(\ref{eq:Laplaces_equation}) implies Dirac's equation.

\section{Cartan's Spinors}
\label{sec:Cartan}
 The concept of Spinors was introduced by Eli
Cartan in 1913\cite{ref:Cartan}. In Cartan's original formulation
spinors were motivated by studying isotropic vectors which are
vectors of zero length. In three dimensions the equation of an
isotropic vector is
\begin{equation}
\label{eq:isotropic_vector}
 x_1^2 + x_2^2 + x_3^2=0
\end{equation}
for complex quantities $x_i$.  A closed form solution to this
equation is realized as
\begin{equation}
\label{eq:Cartan_spinor_solution}
\begin{array}{lccr}
{\displaystyle x_1 =\xi_0^2-\xi_1^2,\ } & {\displaystyle
x_2=i(\xi_0^2+\xi_1^2),}&\ \mathrm{and}\ & {\displaystyle
x_3=-2\xi_0\xi_1}
\end{array}
\end{equation}
 where the two quantities $\xi_i$ are
\[
\begin{array}{lcr}
{\displaystyle\xi_0=\pm\sqrt{\frac{x_1-\imath x_2}{2}}}& \
\mathrm{and} \ & {\displaystyle \xi_1=\pm\sqrt{\frac{-x_1-\imath
x_2}{2}}}
\end{array}.
\]
That the two component object $\xi=(\xi_0,\xi_1)$ is a
spinor\cite{ref:Cartan} can be seen by considering a rotation on the
quantities $v_1=x_1-\imath x_2$, and $v_2=-x_1-\imath x_2$.  If
$v_i$ is rotated by an angle $\alpha$,

 \[v_i\rightarrow v_i \exp(\imath\alpha)
 \]
  then the spinor
component $\xi_0$ is rotated by $\alpha/2$. It is clear that the
spinor is not periodic in $2\pi$ but in $4\pi$.  A quantity of this
type is a spinor and any equation of the form
(\ref{eq:isotropic_vector}) has a spinor solution.

Laplace's equation
\[\left(\frac{\partial^2}{\partial x_1^2}+
\frac{\partial^2}{\partial x_2^2} + \frac{\partial^2}{\partial
x_3^2}\right)\phi=0
\]
can be viewed as an isotropic vector in the following way.  The
components of the vector are the partial derivative operators
$\partial/\partial x_i$ acting on the quantity $\phi$.  As long as
the partial derivatives are restricted to acting on the scalar field
$\phi$ it has a spinor solution given by
\begin{equation}
\label{eq:spinor0}
 \hat{\xi}_0^2=\frac{1}{2}\left(\frac{\partial}{\partial
x_1}-i\frac{\partial}{\partial x_2}\right)=\frac{\partial}{\partial
z_0}
\end{equation}
and
\begin{equation}
\label{eq:spinor1}
\hat{\xi}_1^2=-\frac{1}{2}\left(\frac{\partial}{\partial
x_1}+i\frac{\partial}{\partial x_2}\right)=\frac{\partial}{\partial
z_1}
\end{equation}
where
\[
\begin{array}{lcr}
 {\displaystyle z_0=x_1+ix_2}& \ \mathrm{and}\ & {\displaystyle
 z_1=-x_1+ix_2}
\end{array}
\]and the "hat" notation indicates that the quantities
$\hat{\xi}$ are operators.  The equations
\[
\hat{\xi}_0^2=\frac{\partial}{\partial z_0}
\]
and
\[
\hat{\xi}_1^2=\frac{\partial}{\partial z_1}.
\]
are equations of fractional derivatives of order $1/2$ denoted
$\hat{\xi}_0=D^{1/2}_{z_0}$ and $\hat{\xi}_1=D^{1/2}_{z_1}$.
Fractional derivatives have the property that\cite{ref:Miller_Ross}
\[
D^{1/2}_{z}D^{1/2}_{z}=\frac{\partial}{\partial z}
\]
 and solutions for these fractional derivatives can be
 written\cite{ref:Miller_Ross}
\begin{equation}
D^{\frac{1}{2}}_z \phi=\frac{1}{\Gamma
\left(\frac{1}{2}\right)}\frac{\partial}{\partial z}\int^z_0
(z-t)^{-\frac{1}{2}}\phi(t)dt
\end{equation}
The exact form for these fractional derivatives however, is not
important here.  The important thing to note is that a solution to
Laplace's equation can be written in terms of spinors which are
fractional derivatives.
%\subsection{Spinor Properties}

If we assume that the fractional derivatives $\hat{\xi}_0$ and
$\hat{\xi}_1$ commute then we also have
\begin{eqnarray*}
(\hat{\xi}_0\hat{\xi}_1)^2&=&\hat{\xi}_0\hat{\xi}_0\hat{\xi}_1\hat{\xi}_1 \nonumber \\
&=&-\frac{\partial}{\partial z_0}\frac{\partial}{\partial z_1}\nonumber \\
&=&-\frac{1}{4}\left(\frac{\partial}{\partial x_3}-
\imath\frac{\partial}{\partial x_2}\right)
\left(\frac{\partial}{\partial x_3}+ \imath\frac{\partial}{\partial
x_2}\right)\nonumber \\
&=&-\frac{1}{4}\left(\frac{\partial ^2}{\partial x_2^2}+
\frac{\partial ^2}{\partial x_3^2}\right)\nonumber \\
&=&\frac{1}{4}\frac{\partial^2}{\partial x_1^2}\nonumber\\
\end {eqnarray*}

Using this result combined with Equations~(\ref{eq:spinor0}) and
(\ref{eq:spinor1}) we may write for the components of our vector
\begin{equation}
\label{eq:derivative1_solution}
 \frac{\partial}{\partial x_1}=
-2\hat{\xi}_0\hat{\xi}_1
\end{equation}
\begin{equation}
\label{eq:derivative2_solution} \frac{\partial}{\partial
x_2}=\imath(\hat{\xi}_0^2 + \hat{\xi}_1^2)
\end{equation}
and
\begin{equation}
\label{eq:derivative3_solution}
 \frac{\partial}{\partial x_3}=\hat{\xi}_0^2 - \hat{\xi}_1^2.
\end{equation}

This result gives the explicit solution of our vector quantities
$\frac{\partial}{\partial x_i}$ in terms of the spinor quantities
$\xi_i$.

%Upon complex conjugation we observe that
%\[
%\begin{array}{lcr}
%\left(\hat{\xi}_0^2\right)^*=-\hat{\xi}_1^2 &\mathrm{ and }
%&\left(\hat{\xi}_1^2\right)^*=-\hat{\xi}_0^2.
%\end{array}
%\]
%which implies
%\begin{equation}
%\label{eq:complex_conjugate_spinor_components}
%\begin{array}{lcr}
% \hat{\xi}^*_0=\imath\hat{\xi}_1
%&\mathrm{and} &\hat{\xi}^*_1=\imath\hat{\xi}_0
%\end{array}.
%\end{equation}
%Under complex conjugation therefore our vector becomes\vspace{15pt}
%\[
%\vspace{15pt}
%\begin{array}{lccr}
% {\displaystyle\left(\frac{\partial}{\partial x_1}\right)^*= -\frac{\partial}{\partial
% x_1},}& \ \ \ \
%{\displaystyle\left(\frac{\partial}{\partial
%x_2}\right)^*=\frac{\partial}{\partial x_2}}& \ \ \mathrm{and} \ \ \
%\  &
% {\displaystyle\left(\frac{\partial}{\partial
%x_3}\right)^*=\frac{\partial}{\partial x_3}}.
%\end{array}
%\]
% We see
%therefore that our solution is relevant for an elastic medium
%embedded in a pseudo-Euclidean space in which the vector
%$\partial/\partial x_1$ is pure imaginary. If we write
%\[
%\frac{\partial}{\partial x_1}=\imath\frac{\partial}{\partial t}
%\]
%then Laplace's equation becomes the wave equation
%\[\left(-\frac{\partial^2}{\partial t^2}+
%\frac{\partial^2}{\partial x_2^2} + \frac{\partial^2}{\partial
%x_3^2}\right)\phi=0
%\]

\subsection{Matrix Form}

It can be readily verified that our spinors satisfy the following
equations
\begin{eqnarray*}
\left[\hat{\xi}_0 \frac{\partial}{\partial x_1}+ \hat{\xi}_1
\left(\frac{\partial}{\partial x_3}-i\frac{\partial}{\partial
x_2}\right)\right]\phi=0\\
 \left[\hat{\xi}_0\left(\frac{\partial}{\partial x_3} +
i\frac{\partial}{\partial
x_2}\right)-\hat{\xi}_1\frac{\partial}{\partial x_1}\right]\phi=0
\end{eqnarray*}
and in matrix form
\begin{equation}
\label{eq:dirac_matrix}
 \left(
  \begin{array}{lr}
{\displaystyle\frac{\partial}{\partial x_1}} &
{\displaystyle\frac{\partial}{\partial
x_3}-i\frac{\partial}{\partial
 x_2}} \\[15pt]
{\displaystyle\frac{\partial}{\partial
x_3}+i\frac{\partial}{\partial
 x_2}} & {\displaystyle -\frac{\partial}{\partial x_1}}
  \end{array}
   \right)
 \left(\begin {array}{c}
 {\displaystyle\hat{\xi}_0 }\\[20pt]
  {\displaystyle\hat{\xi}_1}
  \end{array}
  \right)  \phi=0
\end{equation}

The matrix
\[
 X=\left(\begin{array}{lr}
 {\displaystyle\frac{\partial}{\partial x_1}} & {\displaystyle\frac{\partial}{\partial x_3}-i\frac{\partial}{\partial
 x_2}} \\[15pt]
{\displaystyle\frac{\partial}{\partial
x_3}+i\frac{\partial}{\partial
 x_2}} & {\displaystyle -\frac{\partial}{\partial x_1}}
 \end{array}
 \right)
 \]
 is equal to the dot product of the vector $\partial_\mu\equiv\partial/\partial x_\mu$ with the Pauli spin matrices
 \[
 X=\frac{\partial}{\partial x_1}\gamma^1 + \frac{\partial}{\partial
 x_2}\gamma^2 + \frac{\partial}{\partial x_3}\gamma^3
 \]
 where
 \[
 \begin{array}{ccc}
 \gamma_1=\left(\begin{array}{ll}
 1 & 0\\
 0 & -1
 \end{array}
 \right),&
\gamma_2=\left(\begin{array}{ll}
 0 & -i\\
 i & 0
 \end{array}
  \right),&
 \gamma_3=\left(\begin{array}{ll}
 0 & 1\\
 1 & 0
 \end{array}
 \right)
\end{array}
 \] are the Pauli matrices.

So Equation~(\ref{eq:dirac_matrix}) can be written
\begin{equation}
\label{eq:dirac_unstrained}
\sum_{\mu=1}^3\partial_\mu\gamma^\mu\xi=0.
\end{equation}
where we have used the notation $\xi\equiv \hat{\xi}\phi$. This
equation has the form of Dirac's equation in 3 dimensions.  It
describes a spin $1/2$ particle of zero mass that is free of
interactions.
\subsection{Relation to the Dirac Decomposition}
The fact that Laplace' equation and Dirac's equation are related is
not new.  However the decomposition used here is not the same as
that used by Dirac.  In the usual method, starting with the dirac
equation $(i\gamma^\mu\partial_\mu)\Psi=0$ and operating with
$-\imath\gamma^\mu\partial_\mu$ yields Laplace's equation for each
component of the spinor field.  In other words this method results
in not one Laplace's equation but several (one for each component of
the spinor).  Conversely if one starts with Laplace's equation and
tries to recover Dirac's equation one must start with $2$
independent scalars ( $4$ in the usual $4$ dimensional case) in
order to derive the two component spinor equation
(\ref{eq:dirac_unstrained}).

What has been demonstrated in the preceding sections is that
starting with only one scalar quantity satisfying Laplace's equation
Dirac's equation for a two component spinor may be derived.
Furthermore any medium (such as an elastic solid) that has a single
scalar that satisfies Laplace's equation must have a spinor that
satisfies Dirac's equation and such a derivation necessitates the
use of fractional derivatives.

The form of Equation~(\ref{eq:dirac_unstrained}) is relevant for a
massless, non-interacting spin 1/2 particle.  We will now
demonstrate that if one of our internal coordinates is taken to be
periodic a mass term as well as gravitational and electromagnetic
interaction terms appear in Dirac's equation.

\section{Transformation to Internal Coordinates}
\label{sec:internal_coordinates} In section
\ref{sec:fourier_transform} we will take the $x_3^\prime$ coordinate
to be periodic and we will derive equations for the Fourier
components of our fields.  Since the elastic solid is assumed to be
periodic in the internal coordinates we need to translate our
equations of motion from fixed space coordinates to internal
coordinates. Using Equation~(\ref{eq:coordinate_change}) we can
rewrite Equation~(\ref{eq:dirac_unstrained}), as

\begin{equation}
\label{eq:dirac_before_FT}
\sum_{\mu=1}^3\gamma^\mu\left(\partial_\mu'+\sum_\nu\frac{\partial
u_\nu}{\partial x_\mu} \partial_\nu'\right)\xi=0
\end{equation}
or
\[
\sum_{\mu=1}^3\gamma'^\mu\partial'_\mu\xi=0
\]

where $\partial'_\mu=\partial/\partial x'_\mu$ and $\gamma'^\mu$ is
given by
\begin{equation}
\label{eq:modified_gamma_matrices}
\gamma'^\mu=\gamma^\mu+\sum_\alpha\frac{\partial u_\mu}{\partial
x_\alpha}\gamma^\alpha.
\end{equation}

The anticommutator of these matrices is
\begin{eqnarray*}
\{\gamma'^\mu,\gamma'^\nu\}&=&
\{\gamma^\mu+\sum_\alpha\frac{\partial u_\mu}{\partial
x_\alpha}\gamma^\alpha,\gamma^\nu+\sum_\beta\frac{\partial
u_\mu}{\partial x_\beta}\gamma^\beta\}\\
&=&\{\gamma^\mu,\gamma^\nu\}+\sum_\beta
u_{\nu\beta}\{\gamma^\mu,\gamma^\beta\} + \sum_\alpha
u_{\mu\alpha}\{\gamma^\alpha,\gamma^\nu\}+
\sum_{\alpha\beta}u_{\mu\alpha}u_{\nu\beta}\{
\gamma^\alpha,\gamma^\beta\}\\
&=&\delta_{\mu\nu}+\sum\beta u_{\nu\beta}\delta_{\mu\beta}+
\sum_\alpha u_{\mu\alpha}\delta_{\alpha\nu}+ \sum_\alpha\sum_\beta
u_{\mu\alpha}u_{\nu_\beta} \delta_{\alpha\beta}\\
&=&\delta_{\mu\nu}+2u_{\mu\nu}+\sum_\alpha
u_{\mu\alpha}u_{\nu\alpha}\\
&\equiv& g^{\mu\nu}
\end{eqnarray*}

This shows that the gamma matrices have the form of the usual dirac
matrices in a curved space\cite{ref:Brill_Wheeler}. To further
develop the form of Equation~(\ref{eq:dirac_before_FT}) we have to
transform the spinor properties of $\xi$.  As currently written
$\xi$ is a spinor with respect to the $x_i$ coordinates not the
$x'_i$ coordinates. To transform its spinor properties we use a
similarity transformation and write $\xi=S\tilde{\xi}$ where $S$ is
a similarity transformation that takes our spinor in $x_\mu$ to a
spinor in $x'_\mu$.  We will not attempt to give an explicit form
for $S$.  We simply assume (similar to
reference[\cite{ref:Brill_Wheeler}]) that this transformation can be
effected by a real similarity transformation.

We then have
\[
\partial'_\mu\xi=(\partial'_\mu S)\tilde{\xi}+S\partial'_\mu\tilde{\xi}.
\]
Equation~(\ref{eq:dirac_before_FT}) then becomes
\[
\begin{array}{lcl}
 \gamma'_\mu
[S\partial'_\mu\tilde{\xi}+(\partial'_\mu S)\tilde{\xi}]&=&0\\
 \mbox{}
&=&\gamma'_\mu S[\partial'_\mu\tilde{\xi}+S^{-1}(\partial'_\mu
S)\tilde{\xi}]\\
\mbox{} &=&S^{-1}\gamma'_\mu
S[\partial'_\mu\tilde{\xi}+S^{-1}(\partial'_\mu S)\tilde{\xi}]
\end{array}
\]
Using $(\partial'_\mu S^{-1}) S=-S^{-1}(\partial'_\mu S)$. This can
finally be written as
\begin{equation}
\label{eq:dirac_curved_space} \tilde{\gamma}_\mu
[\partial'_\mu-\Gamma_\mu]\tilde{\xi}=0
\end{equation}
where $\Gamma_\mu=(\partial'_\mu S^{-1})S$ and
$\tilde{\gamma}_\mu=S^{-1}\gamma'_\mu S$.
Equation~(\ref{eq:dirac_curved_space}) has the form of the
Einstein-Dirac equation in 3 dimensions for a free particle of zero
mass. The quantity $\partial'_\mu-\Gamma_\mu$ is the covariant
derivative for an object with spin in a curved
space\cite{ref:Brill_Wheeler}. In order to make this identification,
the field $\Gamma_\mu$ must satisfy the additional
equation\cite{ref:Brill_Wheeler,ref:Brill_Cohen}
\begin{equation}
\label{eq:auxiliary_equation}
 \frac{\partial
\tilde{\gamma}^\mu}{\partial
x'^\nu}+\tilde{\gamma}^\beta\Gamma^\mu_{\beta\nu}-\Gamma_\nu\tilde{\gamma}^\mu+\tilde{\gamma}^\mu\Gamma_\nu=0
\end{equation}
 where $\Gamma^\mu_{\beta\nu}$ is the usual Christoffel symbol.
We will now show that this equation does hold for this form of
$\Gamma$.
\subsection{Spin Connection}
To show that Equation~(\ref{eq:auxiliary_equation}) holds we
consider the equation $\partial_\nu\vec{\gamma}=0$ where the vector
$\vec{\gamma}$ is
\[
\vec{\gamma}=\sum_{\mu=1}^3\gamma^\mu \vec{e_\mu}
\]
and $\vec{e_\mu}$ is a unit vector in the $x_\mu$ direction. Since
$\vec{\gamma}$ is a vector, then the quantity
$\partial_\nu\vec{\gamma}=0$ is a tensor equation.  Therefore, in
the primed coordinate system we can immediately write
\[
\sum_{\mu=1}^3 \left(\partial'_\nu
\gamma'^\mu+\gamma'^\beta\Gamma^\mu_{\beta\nu}\right)\vec{e_\mu}=0
\]
where $\gamma'^\mu=\gamma^\mu+\sum_\alpha\frac{\partial
u_\mu}{\partial x_\alpha}\gamma^\alpha$ is the expression of
$\gamma^\mu$ in the primed coordinate system.  Using
$\gamma'^\mu=S\tilde{\gamma}^\mu S^{-1}$, we have

\[
\sum_{\mu=1}^3 \left(\partial'_\nu (S\tilde{\gamma}^\mu
S^{-1})+(S\tilde{\gamma}^\beta
S^{-1})\Gamma^\mu_{\beta\nu}\right)\vec{e_\mu} =0
\]
or
\[\sum_{\mu=1}^3 \left((\partial'_\nu
S)\tilde{\gamma}^\mu S^{-1}+ S(\partial'_\nu \tilde{\gamma}^\mu) S^{-1}+ S
\tilde{\gamma}^\mu (\partial'_\nu S^{-1})+(S\tilde{\gamma}^\beta
S^{-1})\Gamma^\mu_{\beta\nu}\right)\vec{e_\mu}=0.
\]

Multiplying by $S^{-1}$ on the left and $S$ on the right yields
\[
\sum_{\mu=1}^3 \left(S^{-1}(\partial'_\nu S)\tilde{\gamma}^\mu +
(\partial'_\nu \tilde{\gamma}^\mu) +  \tilde{\gamma}^\mu (\partial'_\nu
S^{-1})S+\tilde{\gamma}^\beta \Gamma^\mu_{\beta\nu}\right)\vec{e_\mu} =0
\]

Finally, using $\Gamma_\mu=(\partial_\mu S^{-1})S$ and again noting
that $\partial_\mu S^{-1}S=-S^{-1}\partial_\mu S$ we have,
\[
\tilde{\gamma}^\mu\Gamma_\mu-\Gamma_\mu\tilde{\gamma}^\mu +\left(\partial'_\nu
\tilde{\gamma}^\mu  +\tilde{\gamma}^\beta \Gamma^\mu_{\beta\nu}\right)=0
\]

We have just demonstrated that in the internal coordinates, the
equations of motion of an elastic medium have the same form as the
free-field Einstein Dirac equation for a massless particle in three
dimensions.
\subsection{Physical Content}
Thus far all of the transormations that have been obtained are
"trivial" in the sense that they only result due to changing
coordinates from the unprimed coordinates $x_\mu$ to the primed
coordinates $x_\mu'$.  Changes of coordinates of course do not
result in any new physical content.  In particular the metric
derived in Equation~(\ref{eq:metric}) does not lead to a curved
space.  The Riemann curvature tensor calculated from
Equation~(\ref{eq:metric}) is identically zero.  Likewise the spin
connection $\Gamma_\mu$ is due solely to a gauge transformation
$\xi\rightarrow S\xi'$ and as such contains no physical content
since it can be removed by transforming $\xi'\rightarrow S^{-1}\xi$.

What we will demonstrate in the following sections is that for a
system where one coordinate is periodic, the resulting $2$
dimensional quantities are NOT trivial.  In other words the metric
that determines the dynamics of the Fourier components of $\xi$ does
in fact lead to a curved space and the spin connection cannot be
removed by a gauge transformation.  Furthermore the introduction of
the fourier components will generate extra terms in
Equation~(\ref{eq:dirac_before_FT}) that imply a series of equations
relevant for particles with mass coupled to fields that can be
associated with electromagnetism. We will show that in the low
energy approximation (ie a system in which only the lowest few modes
are present) the equations of motion are identical in form to
Equation~(\ref{eq:full_dirac}).

\section{Interacting particles with mass}
\label{sec:fourier_transform} In this section we again consider a
three dimensional elastic solid but we take the third internal
dimension to be compact with the topology of a circle. All variables
then become periodic functions of $x'_3$ and can be Fourier
transformed. %In preparation for Fourier Transforming transform the
%spinor and isolate the terms involving $x'_3$ and rewrite
%Equation~(\ref{eq:dirac_before_FT}) as
%
%\begin{eqnarray*}
%0&=&\sum_{\mu=1}^3\partial_\mu\gamma^\mu\hat{\xi}\phi\\
%&=&\sum_{\mu=1}^3\partial_\mu\gamma^\mu S\hat{\xi}'\phi\\
%&=&\sum_{\mu=1}^3\gamma^\mu \left[(\partial_\mu S) + S
%\partial_\mu\right]\hat{\xi}'\phi\\
%&=&\sum_{\mu=1}^2\gamma^\mu\left[(\partial_\mu S) + S
%\partial_\mu\right]\hat{\xi}'\phi + \gamma^3\left[(\partial_3 S) + S
%\partial_3\right]\hat{\xi}'\phi\\
%&=&\sum_{\mu=1}^2\gamma^\mu\left[(\partial_\mu S) + S(
%\partial_\mu' + \sum_{\nu=1}^3 u_{\nu\mu}\partial_\nu')\right]\hat{\xi}'\phi + \gamma^3\left[(\partial_3 S) + S(
%\partial_3' + \sum_{\nu=1}^3 u_{\nu 3}\partial_\nu')\right]\hat{\xi}'\phi\\
%\end{eqnarray*}
%
%

In preparation for Fourier Transforming we isolate the terms
involving $x'_3$ and rewrite Equation~(\ref{eq:dirac_before_FT}) as,

\begin{equation}
\label{eq:dirac_curved_space_separated} \sum_{\mu=1}^2\gamma^\mu
\left(\partial_\mu'+\sum_{\nu=1}^2 \frac{\partial u_\nu}{\partial
x_\mu} \partial_\nu' + \frac{\partial u_3}{\partial x_\mu}
\partial_3' \right)\xi
+ \gamma^3\left(\partial_3'+\sum_{\nu=1}^2 \frac{\partial
u_\nu}{\partial x_3} \partial_\nu' + \frac{\partial u_3}{\partial
x_3}
\partial_3' \right)\xi
\end{equation}
We first transform the partial derivatives of the $u_v$ in equation
(\ref{eq:dirac_curved_space_separated}) to obtain
\[
u_{\nu\mu}\equiv\frac{\partial u_\nu}{\partial
x_\mu}=\sum_ku_{\nu\mu,k}e^{ikx_3'}
\]
where $u_{\nu\mu,k}$ is the $k^{th}$ Fourier mode of $\partial
u_\nu/\partial x_\mu$ and $k=2\pi i/a$ with $a$ the length of the
circle formed by the elastic solid in the $x_3'$ direction and $i$
is an integer. Equation~(\ref{eq:dirac_curved_space_separated}) now
becomes,
\begin{eqnarray*}
\lefteqn{\sum_k
e^{ikx_3'}\left[\sum_{\mu=1}^2\gamma^\mu\left(\partial_\mu'\delta_{k,0}+
\sum_{\nu=1}^2
 u_{\nu\mu,k} \partial_\nu' +
u_{3\mu,k}\partial_3'
 \right)\xi \right.}\hspace{1.5in}\\
 & &
 \mbox{}+\left.\gamma^3\left(\partial_3'\delta_{k,0}+\sum_{\nu=1}^2
 u_{\nu 3,k} \partial_\nu' +
u_{33,k}\partial_3'
 \right)\xi\right]=0
\end{eqnarray*}
Next we transform the spinor (noting that it is periodic in $4\pi
a$),
\[
\xi=\sum_q \xi_{q/2} e^{i\frac{q}{2}x_3'}
\] with $q=2\pi j/a$ and $j$ an integer.
This yields,
\begin{eqnarray*}
\lefteqn{\sum_k\sum_q
e^{ix_3'(k+q/2)}\left[\sum_{\mu=1}^2\gamma^\mu\left(\partial_\mu'\delta_{k,0}+\sum_{\nu=1}^2
 u_{\nu\mu,k} \partial_\nu' +
i(q/2) u_{3\mu,k}
 \right)\xi_q/2\right.}\hspace{1.5in}\\
 & & \left.\mbox{}+
 \gamma^3\left(i(q/2)\delta_{k,0}+\sum_{\nu=1}^2
 u_{\nu 3,k} \partial_\nu' + i(q/2)
u_{33,k}
 \right)\xi_{q/2}\right]=0.
\end{eqnarray*}
 This equation is independently true for each distinct
value of $k+q/2=m/2$ or $2k+q=m$ with $k,q,m$ an integer.  Writing
$q=m-2k$ yields finally,
\begin{eqnarray}
\label{eq:dirac_eq_all_modes} \lefteqn{\sum_k
\left[\sum_{\mu=1}^2\gamma^\mu\left(\partial_\mu'\delta_{k,0}+\sum_{\nu=1}^2
 u_{\nu\mu,k} \partial_\nu' +
i\frac{(m-2k)}{2} u_{3\mu,k}
 \right)\xi_{(m-2k)/2}\right.}\hspace{1.5in}\\
 & & \left.\mbox{}+
 \gamma^3\left(i\frac{(m-2k)}{2}\delta_{k,0}+\sum_{\nu=1}^2
 u_{\nu 3,k} \partial_\nu' + i\frac{(m-2k)}{2}
u_{33,k}
 \right)\xi_{(m-2k)/2}\right]=0 \nonumber
\end{eqnarray}
This is an infinite series of equations describing the dynamics of
the fields $\xi_m$.  This set of equations describes the dynamics of
our elastic solid and contains the same information as Laplace's
equation.

So far no approximations have been made. In the next section we will
demonstrate that if only the lowest modes are present, this reduces
to an equation that is identical in form to
Equation~(\ref{eq:full_dirac}).

\subsection{Spectrum of Lowest modes}
\label{sec:lowest_modes}
 We now consider a theory in which only the lowest few modes in
Equation~(\ref{eq:dirac_eq_all_modes}) are present. We therefore
keep only the $m=0,\pm1/2$ modes we obtain the following 3
equations,
\begin{equation}
\label{eq:mode_0}
\sum_{\mu=1}^2\gamma^\mu\left(\partial'_\mu+\sum_{\nu=1}^2u_{\nu\mu,0}\partial_\nu'\right)\xi_0
+ \sum_{\nu=1}^2u_{\nu3,0}\partial'_\nu\gamma^3\xi_0 =0
%\hspace{1.5in} \nonumber\\ &&\hbox{}+ \gamma^3\sum_{\nu=1}^2
%u_{\nu 3,1}\partial_\nu' \xi_{-1} + \gamma^3\sum_{\nu=1}^2 u_{\nu
%3,-1}\partial_\nu' \xi_{1} =0
\end{equation}
\begin{eqnarray}
\label{eq:mode_1}
\lefteqn{\sum_{\mu=1}^2\gamma^\mu\left(\partial'_\mu+\sum_{\nu=1}^2u_{\nu\mu,0}\partial_\nu'+im_{1/2}u_{3\mu,0}\right)\xi_{1/2}
+ \gamma^3\imath m_{1/2}(1+u_{33,0})\xi_{1/2}}\hspace{1.0in} \nonumber \\
 & & \hbox{}+ \gamma^3\sum_{\nu=1}^2u_{\nu3,0}\partial'_\nu\xi_{1/2} +
 \gamma^3\sum_{\nu=1}^2 u_{\nu 3,1}\partial'_\nu \xi_{-1/2}=0
\end{eqnarray}
\begin{eqnarray}
\label{eq:mode_-1}
\lefteqn{\sum_{\mu=1}^2\gamma^\mu\left(\partial'_\mu+\sum_{\nu=1}^2u_{\nu\mu,0}\partial_\nu'+\imath
m_{-1/2}u_{3\mu,0}\right)\xi_{-1/2}
+ \gamma^3\imath m_{-1/2}(1+u_{33,0})\xi_{-1/2}}\hspace{1.0in}\nonumber\\
 & & \hbox{} +
\gamma^3\sum_{\nu=1}^2u_{\nu3,0}\partial'_\nu\xi_{-1/2} +
 \gamma^3\sum_{\nu=1}^2 u_{\nu 3,-1}\partial'_\nu \xi_{1/2}=0
\end{eqnarray}
where $m_i=2\pi i/a$ denotes the Fourier mode with $i$ a half
integer. These equations describe the dynamics of three fields
$\xi_0$ and the coupled fields $\xi_{1/2}$ and $\xi_{-1/2}$.
%The last
%three terms in Equation~(\ref{eq:mode_0}) and the last two terms in
%(\ref{eq:mode_1}) and (\ref{eq:mode_-1}) can be interpreted as
%interaction terms among these fields and all other fields in the
%infinite series. We will not consider interaction terms in this work
%and will therefore drop these terms in the remainder of this text
%and focus on the uncoupled dynamics of each equation.
The first equation ($m=0$ mode) describes the dynamics of a
massless, free particle.  We will not attempt to identify this mode
with any physical particle but we simply note that in this
approximation this equation is completely uncoupled from the
$m=\pm1/2$ modes and therefore its dynamics are independent and have
no affect on these other modes.

We now examine the equations describing $\xi_{1/2}$ and
$\xi_{-1/2}$. These two equations can be combined by noting that for
real fields, $u_{\mu\nu,k}=u^\ast_{\mu\nu,-k}$. The $m=\pm 1/2$
modes can now be combined into the single equation
\begin{eqnarray}
\label{eq:Psi}
\lefteqn{\sum_{\mu=1}^2\gamma^\mu\left(\partial'_\mu+\sum_{\nu=1}^2u_{\nu\mu,0}\partial_\nu'+im_{1/2}u_{3\mu,0}\right)\Psi
+ \gamma^3\imath m_{1/2}(1+u_{33,0})\Psi}\hspace{1.0in} \nonumber \\
 & & \hbox{}+ \gamma^3\sum_{\nu=1}^2u_{\nu3,0}\partial'_\nu\Psi +
 \gamma^3\sum_{\nu=1}^2 u_{\nu 3,1}\partial'_\nu \Psi^\ast=0,
\end{eqnarray}
where $\Psi=\xi_{1/2}+\xi^\ast_{-1/2}$. To put this equation into a
more recognizable form we multiply Equation~(\ref{eq:Psi}) by
$\gamma^3$ from the left and define
\begin{equation}
\label{eq:dirac_matrices_FT}
\gamma'^\mu=\gamma^3\gamma^\mu+\sum_{\beta=1}^2\gamma^3\gamma^\beta
u_{\mu\beta,0}.
\end{equation}
 These matrices with $\mu=1,2$ and $\nu=1,2$ satisfy the
anticommutation relations
\[
\left\{\gamma'^\mu,\gamma'^\mu\right\}= \delta_{\mu\nu} +
(u_{\mu\nu,0}+u_{\nu\mu,0})+\sum_{\beta=1}^2u_{\mu\beta,0}u_{\nu\beta,0}.
\]
If we insist that our new matrices satisfy
$\left\{\gamma^\mu,\gamma^\nu\right\}=g^{\mu\nu}$ then we are led to
define
\begin{equation}
\label{eq:commutation_relations_FT} g^{\mu\nu}\equiv\delta_{\mu\nu}
+
(u_{\mu\nu,0}+u_{\nu\mu,0})+\sum_{\beta=1}^2u_{\mu\beta,0}u_{\nu\beta,0}.
\end{equation}  This is the metric for our two dimensional subspace and
it does not have the form of a simple coordinate transformation on a
flat space metric like that of section \ref{sec:elasticity_theory}.

Equation~(\ref{eq:Psi}) can now be rewritten as
\begin{eqnarray}
\label{eq:dirac_recognizable_form}
\lefteqn{\sum_{\mu=1}^2\gamma'^\mu\left(\partial'_\mu+\imath
m_{1/2}u_{3\mu,0}\right)\Psi -\imath
m_{1/2}u_{3\mu,0}\sum_{\beta=1}^2\gamma^3\gamma^\beta
u_{\mu\beta,0}\Psi+\imath m_{1/2}(1+u_{33,0})\xi_{1/2}}\hspace{2.5in} \nonumber \\
& & \hbox{}+ \sum_{\nu=1}^2u_{\nu3,0}\partial'_\nu\Psi
+\sum_{\nu=1}^2 u_{\nu 3,1}\partial'_\nu \Psi^\ast=0.
\end{eqnarray}

As we did for Equation~(\ref{eq:dirac_curved_space_separated}), in
going from the $x$ to the $x'$ coordinates, we assume that the
spinor properties of $\xi_{1/2}$ may be transformed using a real
similarity transformation and writing
$\xi_{1/2}=S\tilde{\xi}_{1/2}$.  Transforming $\Psi$ in this way and
multiplying on the left by $S^{-1}$ gives us the following form for
the $m=\pm1/2$ modes,
\begin{eqnarray}
\label{eq:dirac_final_form} \lefteqn{\sum_{\mu=1}^2
S^{-1}\gamma'^\mu S \left(\partial'_\mu+\imath m_{1/2}u_{3\mu,0}+
S^{-1}\partial'_\mu S\right)\tilde{\Psi} -\imath
m_{1/2}u_{3\mu,0}\sum_{\beta=1}^2 S^{-1}\gamma^\beta S
u_{\mu\beta,0}\tilde{\Psi}} \hspace{.1in}\\
 & & \mbox{}+  \imath
m_{1/2}(1+u_{33,0})\tilde{\Psi}+ \sum_{\nu=1}^2
u_{\nu3,0}(\partial'_\nu +S^{-1}\partial'_\nu S)\tilde{\Psi}
+\sum_{\nu=1}^2 u_{\nu 3,1}(\partial'_\nu+S^{-1}\partial'_\nu S)
\Psi^\ast\nonumber
\end{eqnarray}

We now examine each quantity in
Equation~(\ref{eq:dirac_final_form}). As before, we identify
$\tilde{\gamma}^\mu=S^{-1}\gamma'^\mu S$ with the transformed gamma
matrix and $\Gamma_\mu=(\partial'_\mu S^{-1})S$ with the spin
connection. We also would like to identify the quantity
$A_\mu=m_1u_{3\mu,0}$ in the first term with the electromagnetic
potential and the third term in Equation~(\ref{eq:dirac_final_form})
as a mass term with $m=m_1(1+u_{33,0})$ which implies that the field
$u_{33,0}$ provides a mass for our $\Psi$ particle. Let us further
assume that the quantities $u_{\mu\nu}$ are small compared to unity
so that the second term may be neglected as being of order
$u^2_{\mu\nu}$ (ie we are now assuming that our medium undergoes
only small deformations).  If these identifications are made we can
write Equation~(\ref{eq:dirac_final_form}) in the final form
\begin{eqnarray}
\label{eq:dirac_final_form2} \lefteqn{\sum_{\mu=1}^2
\tilde{\gamma}^\mu \left(\imath\partial'_\mu+ \imath\Gamma_\mu-
A_\mu\right)\tilde{\Psi} -
m\tilde{\Psi} } \hspace{1.5in}\\
 & & \mbox{}+ \imath\sum_{\nu=1}^2u_{\nu3,0}(\partial'_\nu-\Gamma_\nu)\tilde{\Psi}
 +
\imath\sum_{\nu=1}^2 u_{\nu 3,1}(\partial'_\nu-\Gamma_\nu)
\Psi^\ast\nonumber.
\end{eqnarray}  Notice the formal similarity of this equation to
Equation~(\ref{eq:full_dirac}).  The first two terms have exactly
the form of Dirac's equation for a spin $1/2$ particle of mass $m$
in curved space interacting with the electromagnetic vector
potential $A_\mu$. Note that the mass term and the electromagnetic
potentials were not added by hand but emerged naturally from the
formalism. The nature of the last two terms in
Equation~(\ref{eq:dirac_final_form2}) are unknown. They don't appear
in the usual statement of Dirac's equation and their implications
are unknown.

Equation~(\ref{eq:dirac_final_form2}) is the central result of this
work.  We have as yet not shown that the dynamics of the fields
$A_\mu$ are consistent with their identification as the
electromagnetic vector potential.  To truly claim that the quantity
$u_{\mu 3,0}$ is the electromagnetic potential it must be shown to
satisfy Maxwell's equations.  We believe however that the formal
correspondence between Equation~(\ref{eq:dirac_final_form2}) and
Dirac's equation is significant in its own right and will not, in
this paper, pursue the question of whether Maxwell's equation or the
Einstein Field equations are satisfied. Before concluding we note
that although our derivation assumed that we were working in three
dimensional space, the formalism extends to any number of
dimensions\cite{ref:Cartan,ref:Brauer_Weyl}.  The major difference
is that in the three dimensional case, we were able to find an
explicit solution for the components of a spinor in terms the
components of the vector $\partial _\mu$.  An explicit solution
might not exist in general. Nevertheless it can be
shown\cite{ref:Cartan} that the quadratic form in Laplace's equation
implies the existence of a multicomponent spinor $\xi$ satisfying a
dirac-like equation in any dimension.

\section{Conclusions}
We have taken a model of an elastic medium and derived an equation
of motion that has the same form as Dirac's equation in the presence
of electromagnetism and gravity.  We derived our equation by using
the formalism of Cartan to reduce the quadratic form of Laplace's
equation to the linear form of Dirac's equation. We further assumed
that one coordinate was compact and upon Fourier transforming this
coordinate we obtained, in a natural way, a mass term and an
electromagnetic interaction term in the equations of motion.

\end{document}